\documentstyle[12pt]{article}
\begin{document}
\def \beq{\begin{equation}}
\def \eeq{\end{equation}}
\def \bea{\begin{eqnarray}}
\def \eea{\end{eqnarray}}
\def \tt {t\overline {t}}
\def \ee {e^+e^-}
\def \ra {\rightarrow}
\def \g {\gamma}
\def \cvg {c_v^{\gamma}}
\def \cag {c_a^{\gamma}}
\def \cvz {c_v^Z}
\def \caz {c_a^Z}
\def \cdg {c_d^{\gamma}}
\def \cdz {c_d^Z}
\def \cdgz {c_d^{\gamma,Z}}
\def \t{$t\;$}
\def \toverline{$\overline t$}
\def \tbar{$\overline t\;$}
\def \el{E_l}
\def \thetal{\theta_l}
\def \phil{\phi_l}
\def \CP{$CP$}
\def \f{\frac}
\def \o{\overline}
\def \n{\noindent}

\begin{flushright}
PRL-TH-95/17\\
hep-ph/9509299
\end{flushright}
\vskip 3mm
\begin{center}
{\huge Decay-lepton angular distribution in polarized $e^+e^- \ra
\tt$ and $CP$-violating dipole couplings of the top quark }
\vskip .5cm
{\large P. Poulose and Saurabh D. Rindani}
\vskip .25cm
{\it Theory Group, Physical Research Laboratory \\ Navrangpura,
Ahmedabad 380009, India}
\vskip 1cm
{\bf Abstract}
\end{center}
\vskip .2cm
In the presence of an electric dipole coupling of $\tt$
to a photon, and an analogous ``weak" dipole coupling to the $Z$,
$CP$ violation in the process $\ee \ra \tt$ leads to the polarization
of the top and anti-top. This polarization can be analyzed by
studying the energy and angular distributions of a decay charged
lepton (anti-lepton) when the top (anti-top) decays leptonically. We
have obtained analytic expressions for these distributions when
either $t$ or $\overline t$ decays leptonically. We study two types
of simple $CP$-violating angular asymmetries which do not need the
full reconstruction of the $t$ or $\o t$. These together can help to
determine the electric and weak dipole form factors independently. We
have also shown how the use of longitudinal beam polarization can
help to do this with only one asymmetry measurement, and to improve
the sensitivity.
\newpage
\section{Introduction}

Experiments at the Tevatron have seen evidence for the top qark with
mass in
the range of about 170-200 GeV \cite{expt}.  Future runs of the
experiment will be able to
determine the mass more precisely and also determine other properties
of the top quark.  $\tt$ pairs will be produced more copiously at
proposed
$e^+ e^-$ linear colliders operating above threshold.  It would then
be
possible to investigate these properties further.

While the standard model (SM) predicts $CP$ violation outside
the \mbox{$K$-,}
$D$- and $B$-meson systems to be unobservably small,
in some extensions of SM, $CP$ violation might be considerably
enhanced, especially in the presence of a heavy top quark.  In
particular,
$CP$-violating electric dipole form factor of the top quark, and
the analogous $CP$-violating ``weak" dipole form factor in the $\tt$
coupling
to $Z$, could be enhanced.  These $CP$-violating form factors could
be
determined in a model-independent way at high energy $\ee$ linear
colliders,
where $\ee \ra \tt$ would proceed through virtual $\g$ and $Z$
exchange.

Since a heavy top quark ($m_t \ge 120$ GeV) is expected to decay
before it
hadronizes \cite{heavytop}, it has been suggested \cite{toppol} that
top
polarization asymmetry in
$\ee \ra \tt$ can be used to determine the $CP$-violating dipole form
factors,
since polarization information would be retained in the decay product
distribution.  Experiments have been proposed in which the
$CP$-violating
dipole couplings could be measured in decay momentum
correlations \cite{bern,atwood,cuypers} or
asymmetries \cite{asymm,PP}, even with beam polarization
\cite{atwood,PP}.  These suggestions on the
measurement of asymmetries have concentrated on experiments requiring
the reconstruction of the top-quark momentum (with the exception of
lepton  energy asymmetry \cite{toppol,asymm,PP}).  In this note we
look at very simple lepton angular asymmetries
which do not require the experimental determination of the $t$ or
$\o{t}$ momentum.  Being single-lepton asymmetries, they do not
require
both $t$ and $\o{t}$ to decay leptonically. Since either $t$ or
$\o{t}$
is also allowed to decay hadronically, there is a gain in statistics.

Our results are based on fully analytical calculation of single
lepton
distributions in the production and subsequent decay of $\tt$. We
present
fully differential distribution as well as the distribution in the
polar angle of the lepton with respect to the beam direction in the
centre-of-mass (cm) frame.
These distributions in the absence of $CP$-violating dipole couplings
were obtained earlier by Arens and Sehgal \cite{arens}, using
the technique of Kawasaki, Shirafuji and
Tsai \cite{tsai}.  We have included $CP$-violating effects in $\tt$
production, and obtained the distributions using the equivalent
helicity-amplitude technique. Our results  agree with the
results of \cite{arens} in the limit of vanishing dipole moments.

We have also included the effect of electron longitudinal
polarization,
likely to be easily available at linear colliders.  In an
earlier paper \cite{PP},
we had shown how polarization helps to put independent limits on
electric
and weak dipole couplings, while providing greater sensitivity in the
case of asymmetries.  We also demonstrate these
advantages for the present case, strengthening the case for
polarization studies.

The rest of the paper is organized as follows. In Sec. II, we
describe the
cal\-cula\-tion of the decay-lepton angular dist\-ribu\-tion from a
decaying  $t$ or $\o t$
in $\ee \ra \tt$. In Sec. III we describe $CP$-violating asymmetries
and  obtain expressions for them. Numerical results are presented in
Sec. IV, and Sec. V contains our conclusions. The Appendix contains
certain  expressions which are too lengthy to be put in the main
text.

\section{Calculation of  lepton  angular distributions}

We describe in this Section the calculation of $l^+\; (l^-)$
distribution in
$\ee \ra \tt$ and the subsequent decay $t \ra b l^+ \nu_l\;
(\overline{t} \ra
\overline{b} l^- \overline{\nu_l})$.  We adopt the narrow-width
approximation for
$t$ and $\overline{t}$, as well as for $W^{\pm}$ produced in
$t,\;\overline{t}$
decay.

  We assume the top quark couplings to $\g$ and $Z$ to be
given by the vertex factor  $ie\Gamma_\mu^j$, where
\beq
\Gamma_\mu^j\;=\;c_v^j\,\g_\mu\;+\;c_a^j\,\g_\mu\,\g_5\;+
\;\f{c_d^j}{2\,m_t}\,i\g_5\,
(p_t\,-\,p_{\overline{t}})_{\mu},\;\;j\;=\;\g,Z,
\eeq
with
\bea
\cvg&=&\f{2}{3},\:\;\;\cag\;=\;0, \nonumber \\
\cdz&=&\f {\left(\f{1}{4}-\f{2}{3} \,x_w\right)}
{\sqrt{x_w\,(1-x_w)}},
 \\
\caz&=&-\f{1}{4\sqrt{x_w\,(1-x_w)}}, \nonumber
\eea
and $x_w=sin^2\theta_w$, $\theta_w$ being the weak mixing
angle.
We have assumed in (1) that the only addition to the SM
couplings $c^{{\g},Z}_{v,a}$ are the $CP$-violating electric and weak
dipole
form factors, $e\cdg/m_t$ and $e\cdz/m_t$, which are
assumed small.  Use has also been made of the Dirac equation in
rewriting the usual dipole coupling
$\sigma_{\mu\nu}(p_t+p_{\overline{t}})^{\nu}\g_5$ as
$i\g_5(p_t-p_{\overline{t}})_{\mu}$, dropping small corrections
to the vector and axial-vector couplings.  We assume that there
is no $CP$ violation in $t$, $\overline{t}$ decay \cite{decay}.

The helicity amplitudes for $\ee \ra \g^*,Z^* \ra \tt$ in the
cm frame, including $\cdgz$ couplings, have been
given in \cite{asymm} (see
also Kane {\it et al.}, ref. \cite{toppol}), so we do not repeat
them here.  The non-vanishing
helicity amplitudes, respectively $M$ and $\overline{M}$, for
\[t \ra b W^+,\;
\;W^+ \ra l^+ \nu_l\]
and\[\overline{t} \ra \overline{b}W^-,\;\; W^- \ra
l^-\overline{\nu_l}\]
in the respective rest frames of $t$, $\overline{t}$, are given
below (we assume standard model couplings and neglect all masses
except $m_t$, the top mass):
\bea
M_{+-+-}\!\!&=&\!\!8g^2\Delta_W(q)\cos\f{\theta_{l^+}}
{2}\left[\cos \f{\theta_{\nu_l}}{2} \sin
\f{\theta_b} {2} e^{-i\phi_b} - \sin \f{\theta_{\nu_l}}{2} \cos
\f{\theta_b} {2} e^{-i\phi_{\nu_l}}\right] e^{i\phi_{l^+}}
,\nonumber \\
&& \\
M_{--+-}\!&=&\!\!8g^2\Delta_W(q)\sin \f{\theta_{l^+}}
{2}\left[\cos \f{\theta_{\nu_l}}{2} \sin
\f{\theta_b} {2} e^{-i\phi_b} - \sin \f{\theta_{\nu_l}}{2} \cos
\f{\theta_b} {2} e^{-i\phi_{\nu_l}}\right],\\
\overline{M}_{++-+}\!\!&=&\!\!8g^2\Delta_W(q)\cos\f{\theta_{l^-}}
{2}\left[\cos \f{\theta_{\overline {\nu_l}}}{2} \sin
\f{\theta_{\overline b}} {2} e^{i\phi_{\overline {\nu_l}}} - \sin
\f{\theta_{\overline {\nu_l}}}{2} \cos
\f{\theta_{\overline b}} {2} e^{i\phi_{\overline b}}\right] , \\
\overline{M}_{-+-+}\!\!&=&\!\!8g^2\Delta_W(q)\sin\f{\theta_{l^-}}
{2}\left[\cos \f{\theta_{\overline {\nu_l}}}{2} \sin
\f{\theta_{\overline b}} {2} e^{i\phi_{\overline {\nu_l}}} - \sin
\f{\theta_{\overline {\nu_l}}}{2} \cos
\f{\theta_{\overline b}} {2} e^{i\phi_{\overline b}}\right]
e^{-i\phi_{l^-}} ,\nonumber \\
&&
\eea
where
\beq
\Delta_W(q)=\f{1}{q^2-m_W^2+i\Gamma_Wm_W}
\eeq
is the $W$ propagator, $q$ being the total $\ee$ momentum.
The subscripts $\pm$ refer to signs of the helicities, the order of
the helicities being $t$, $b$, $l^+$, $\nu_l$ ($\overline{t}$,
$\overline{b}$, $l^-$, $\overline{\nu_l}$).
The various $\theta$'s are polar angles of the particles and
antiparticles labelled by
the suffixes with respect to a $z$ axis in the direction in which the
top momentum is boosted to go to the cm frame. $\phi$'s
are the azimuthal angles with respect to an $x$ axis chosen in the
plane containing the $e^-$ and $t$ directions.

Combining the production and decay amplitudes in the narrow-width
approximation for $t,\overline{t},W^+,W^-$, and using appropriate
Lorentz
boosts to calculate everything in the $\ee$ cm frame, we get the
$l^+$
and $l^-$ distributions for the case of $e^-$, $e^+$ with
polarization
$P_e$, $P_{\o e}$ to be:

\begin{eqnarray}
\lefteqn{\frac{d\sigma^{\pm}}{d\cos\theta_t dE_l d\cos\theta_l
d\phi_l}= \frac{3\alpha^4\beta}{16x_w^2\sqrt{s}}
 \frac{E_l}{\Gamma_t \Gamma_W m_W}
\left(  \frac{1}{1-\beta\cos\theta_{tl}}
  -\frac{4 E_l
}{\sqrt{s}(1-\beta^2)}\right)}\nonumber \\
&&\times \left\{ \left(
A_0+A_1 \cos\theta_t + A_2 \cos^2\theta_t \right) (1-\beta \cos
\theta_{tl})\right. \nonumber \\
&&\left. +\left( B_0^{\pm} + B_1 \cos\theta_t + B_2^{\pm}
\cos^2\theta_t \right) (\cos\theta_{tl}-\beta )\right. \nonumber
\\
&&\left. + \left( C_0^{\pm} + C_1^{\pm} \cos\theta_t \right)
(1-\beta^2) \sin\theta_t \sin\theta_l (\cos\theta_t \cos\phi_l -
\sin\theta_t \cot\theta_l)
 \right.
\nonumber \\
&& \left. + \left( D_0^{\pm} + D_1^{\pm} \cos\theta_t \right)
(1-\beta^2) \sin\theta_t \sin\theta_l \sin\phi_l \right\}.
\end{eqnarray}
The quantities $A_i$, $B_i$, $C_i$ and $D_i$ occurring in the above
equation are functions of the masses, $s$, the degrees of $e$ and $\o
e$  polarization ($P_e$ and $P_{\o e}$), and the coupling constants.
They are listed in the Appendix.

In eq. (8), $\sigma^+$ and $\sigma^-$ refer respectively to $l^+$ and
$l^-$ distributions, with the same notation for the kinematic
variables of
particles and antiparticles.  Thus, $\theta_t$, is the polar angle of
$t$  (or \tbar), and $\el,\;\thetal,\;\phil$ are the energy, polar
angle and azimuthal angle of $l^+$ (or $l^-$).
All the angles are now in the cm frame, with the $z$ axis chosen
along the $e^-$ momentum, and the $x$ axis chosen in the plane
containing the $e^-$ and $t$ directions.
$\theta_{tl}$ is the angle between
the $t$ and $l^+$ directions (or $\overline t$ and $l^-$ directions).
$\beta$
is the $t$ (or $\o t$) velocity: \(\beta=\sqrt{1-4m_t^2/s}\), and
$\gamma = 1/\sqrt{1-\beta^2}$.

Since we are mainly interested in $\thetal$ distributions here, we
first integrate over $\el$ between limits\[ \f{m_W^2}{m_t^2}
\f{\sqrt{s}}{4}
\f{1-\beta^2}{1-\beta\;{\rm cos}\,\theta_{tl}}\; < \; \el \; < \;
\f{\sqrt{s}}{4}
\f{1-\beta^2}{1-\beta\,{\rm cos}\,\theta_{tl}},\]
then over $\phi$ from 0 to $2 \pi$, and finally over cos$\,\theta_t$
from $-1$ to +1.  After some lengthy algebra, we get the final result
as
\bea
\f{d\sigma^{\pm}}{d\cos\theta_l}&=&\frac{3\pi\alpha^2}{32s}B_tB_{\o t}
\left\{4A_0
-2A_1\left(\f{1-\beta^2}{\beta^2} \log\f{1+\beta}{1-\beta}-
\f{2}{\beta}\right) \cos\theta_l \right.\nonumber \\
&&\left. + 2A_2 \left(
\f{1-\beta^2}{\beta^3}\log\f{1+\beta}{1-\beta}
(1-3\cos^2\theta_l) \right. \right. \nonumber \\
&& \left. \left. - \f{2}{\beta^2} (1-3\cos^2\theta_l-\beta^2+2 \beta^2
\cos^2\theta_l) \right) \right. \nonumber \\
&&\left. + 2B_1\f{1-\beta^2}{\beta^2}  \left(
\f{1}{\beta}\log\f{1-\beta}{1-\beta} -
2 \right) \cos\theta_l \right. \nonumber \\
&&  \left. + B_2^{\pm} \f{1}{\beta^3}
\left( \f{\beta^2-2}{\beta} \log\f{1+\beta}{1-\beta} + 6 \right)
(1-3\cos^2\theta_l) \right. \nonumber \\
&&\left. +2C_0^{\pm}\f{1-\beta^2}{\beta^2} \left( \f{1}{\beta}
\log\f{1+\beta}{1-\beta} - 2 \right) \cos\theta_l
\right. \nonumber \\
&& \left. - C_1^{\pm}\f{1}{\beta^3} \left( \f{3(1-\beta^2)}{\beta}
\log\f{1+\beta}{1-\beta} -2(3-2\beta^2)\right) (1-3
\cos^2\theta_l) \right\},\nonumber \\
&&
\eea
where $B_t$ and $B_{\o t}$ are respectively the branching ratios of
$t$ and
$\o t$ into the final states being considered.

We have compared our expression for the angular distribution with the
one in
\cite{arens} in the limit of vanishing dipole moments and vanishing
beam
polarization, and found agreement.

\section{$CP$-violating angular asymmetries}

We define two independent \CP-violating asymmetries, which depend on
different
linear combinations of Im$\cdg$ and Im$\cdz$ . (It is not
possible to define \CP-odd
 quantities which determine Re$\cdgz$ using single-lepton
distributions, as can
be seen from the expression for the \CP-odd combination
$\frac{d\sigma^+}{d\cos\theta_l}(\theta_l)
-\frac{d\sigma^-}{d\cos\theta_l}(\pi-\theta_l)$). One is simply the
total
lepton-charge asymmetry, with a cut-off of $\theta_0$ on the forward
and
backward directions:
\beq
{\cal A}_{ch}(\theta_0)=\frac{
{\displaystyle		\int_{\theta_0}^{\pi-\theta_0}}d\theta_l
{\displaystyle          \left( \frac{d\sigma^+}{d\theta_l}
		-	\frac{d\sigma^-}{d\theta_l}\right)}}
{
{\displaystyle		\int_{\theta_0}^{\pi-\theta_0}}d\theta_l
{\displaystyle          \left( \frac{d\sigma^+}{d\theta_l} +
\frac{d\sigma^-}{d\theta_l}\right)}}.
\eeq
The other is the lept\-onic forward-backward asy\-mmetry com\-bined
with charge asy\-mmetry, again with the angles within $\theta_0$ of
the forward and back\-ward directions excluded:
\beq
{\cal A}_{fb}(\theta_0)= \frac{ {\displaystyle
\int_{\theta_0}^{\frac{\pi}{2}}}d\theta_l {\displaystyle
\left( \frac{d\sigma^+}{d\theta_l} -
\frac{d\sigma^-}{d\theta_l}\right)} {\displaystyle
-\int^{\pi-\theta_0}_{\frac{\pi}{2}}}d\theta_l {\displaystyle
\left( \frac{d\sigma^+}{d\theta_l} -	\frac{d\sigma^-}{d\theta_l}
\right)}}
{
{\displaystyle		\int_{\theta_0}^{\pi-\theta_0}}d\theta_l
{\displaystyle          \left( \frac{d\sigma^+}{d\theta_l} +
\frac{d\sigma^-}{d\theta_l}\right)}}.
\eeq

These asymmetries are a measure of $CP$ violation in the unpolarized
case and in the case when polarization is present, but
$P_e=-P_{\overline{e}}$.  When $P_e\neq -P_{\overline{e}}$, the
initial state is not invariant under $CP$, and therefore
$CP$-invariant interactions can contribute to the asymmetries.
However, to the leading order in $\alpha$, these $CP$-invariant
contributions vanish in the limit $m_e=0$.  Order-$\alpha$ collinear
helicity-flip photon emission can give a $CP$-even contribution.
However, this background can be suppressed by a suitable cut on the
visible energy.

The expressions for ${\cal A}_{ch}(\theta_0)$ and ${\cal
A}_{fb}(\theta_0)$ may be derived from eq. (9) making use of the
expressions in the Appendix, and are given below.
\begin{eqnarray}
\lefteqn{{\cal A}_{ch}(\theta_0)=\frac{1}{2\sigma
(\theta_0)}\frac{3\pi\alpha^2} {4s}B_tB_{\o
t}\,2\cos\theta_0\sin^2\theta_0
\left(
(1-\beta^2)\log\frac{1+\beta}{1-\beta}-2\beta\right)}\nonumber\\
&&\times\left( {\rm Im}\cdg \left\{\left[ 2 c_v^
{\gamma}+(r_L+r_R)c_v^Z\right](1-P_eP_{\overline
e})+(r_L-r_R)c_v^Z(P_{\overline e}-P_e)
\right\} \right.\nonumber \\&&\left.
+ {\rm Im}\cdz \left\{\left[ (r_L+r_R) c_v^
{\gamma}+(r_L^2+r_R^2)c_v^Z\right](1-P_eP_{\overline e})+
\left[(r_L-r_R)c_v^{\gamma}\right.\right.\right.\nonumber\\
&&\left.\left.\left.+(r_L^2-r_R^2)c_v^Z\right] (P_{\overline
e}-P_e)\right\} \right);
\end{eqnarray}
\begin{eqnarray}
{\cal A}_{fb}(\theta_0)&=&\frac{1}{2\sigma
(\theta_0)}\frac{3\pi\alpha^2} {2s}B_tB_{\o t}\,\cos^2\theta_0
\left(
(1-\beta^2)\log\frac{1+\beta}{1-\beta}-2\beta\right)c_a^Z\nonumber\\
&&\times \left\{ {\rm Im}\cdg \left[ (r_L-r_R)(1-P_eP_{\overline
e})+(r_L+r_R)(P_{\overline e}-P_e)
\right] \right.\nonumber \\&&\left.
+ {\rm Im}\cdz \left[ (r_L^2-r_R^2)(1-P_eP_{\overline e})
+(r_L^2+r_R^2)(P_{\overline e}-P_e)\right]\right\}.
\end{eqnarray}
Here $\sigma(\theta_0)$ is the cross section for $l^+$ or $l^-$
production with a cut-off $\theta_0$, and is given by
\begin{eqnarray}
\sigma(\theta_0)&=& \frac{3\pi\alpha^2}{8s}B_tB_{\o t} \,
2\cos\theta_0\left(
\left\{(1-\beta^2)\log\frac{1+\beta}{1-\beta}\sin^2\theta_0
\right. \right. \nonumber \\
&& \left. \left. + 2\beta\left[1+(1-\frac{2}{3}
\beta^2)\cos^2\theta_0\right]\right\} \right.\nonumber\\
&& \times \left.
\left\{\left[2{c_v^{\g}}^2+2c_v^{\g}c_v^Z(r_L+r_R)+{c_v^Z}^2
(r_L^2+r_R^2)\right](1-P_eP_{\overline e})\right.\right.\nonumber \\
&&\left. \left. +c_v^Z\left[(r_L-r_R)c_v^{\g} +
(r_L^2-r_R^2)c_v^Z\right](P_{\overline e}-P_e)\right\}\right.
\nonumber \\
&&+\left.\left\{(1-\beta^2)\log\frac{1+\beta}{1-\beta}\sin^2\theta_0
+ 2\beta\left[2\beta^2-1+(1-\frac{2}{3}
\beta^2)\cos^2\theta_0\right]\right\}\right.\nonumber \\
&&\times\left. {c_a^Z}^2 \left\{(r_L^2+r_R^2)(1-P_eP_{\overline e}) +
(r_L^2-r_R^2) (P_{\overline e}-P_e)\right\} -2(1-\beta^2)\right.
\nonumber \\ &&\left. \times\left( \log\frac{ 1+\beta}{1-\beta} - 2
\right)\sin^2 \theta_0
c_a^Z\left\{\left[(r_L+r_R)c_v^{\g} + (r_L^2+r_R^2) c_v^Z \right]
\right.\right. \nonumber \\ &&\times\left. \left.(1-P_eP_{\overline
e}) + \left[ (r_L-r_R)c_v^{\g}+ (r_L^2-r_R^2) c_v^Z
\right] (P_{\overline e}-P_e)\right\}
\right).
\end{eqnarray}

We note the curious fact that ${\cal A}_{ch}(\theta_0)$ vanishes for
$\theta_0=0$. This implies that the $CP$-violating charge asymmetry
does not exist unless a cut-off is imposed on the lepton production
angle. ${\cal A}_{fb}(\theta_0)$, however, is nonzero for
$\theta_0=0$.

It is also possible to obtain a variety of $CP$-odd correlations
using the analytic form (9). However, we restrict ourselves here to
an analysis of the consequences of ${\cal A}_{ch}$ and ${\cal
A}_{fb}$, without and with beam polarization.

\section{Numerical Results}

In this section we describe the numerical results for the calculation
of 90\% confidence level (CL) limits that could be put on Im$\cdgz$
using the asymmetries described in the previous sections, as well as
the $CP$-odd part of the angular distribution in eq. (9).

We look at only semileptonic final states. That is to say, when $t$
decays leptonically, we assume $\o t$ decays hadronically, and {\it
vice versa}. We sum over the electron and muon decay channels. Thus,
$B_tB_{\o t}$ is taken to be $2/3\times2/9$.  The number of events
for various relevant $\theta_0$ and for beam polarizations $P_e=0$,
$\pm 0.5$ are listed in Table 1.

In each case we have derived simultaneous 90\% CL limits on
Im$c_d^{\g}$ and Im$c_d^Z$ that could be put in an experiment at a
future linear colider with $\sqrt{s}=500$ GeV and an integrated
luminosity of 10 fb$^{-1}$. We do this by equating the asymmetry
(${\cal A}_{ch}$ or ${\cal A}_{fb}$) to $2.15/\sqrt{N}$, where $N$ is
the total number of expected events. In the unpolarized case, each of
${\cal A}_{ch}$ and ${\cal A}_{fb}$ gives a band of allowed values in
the Im$c_d^{\g}-$Im$c_d^Z$ plane. If both ${\cal A}_{ch}$ and ${\cal
A}_{fb}$ are looked for in an experiment, the intersection region of
the corresponding bands determines the best 90\% CL limits which can
be put simultaneously on Im$c_d^{\g}$ and Im$c_d^Z$. These best
results are obtained for $\theta_0=35^\circ$ and are shown in Fig.
1(a) and Fig. 1(b), for two values of the top mass, $m_t=174$ GeV,
and $m_t=200$ GeV respectively.

We see  from Fig. 1 that the 90\% CL limits that could be put on
Im$\cdg$ and Im$\cdz$ simultaneously are, respectively, 2.4 and 17,
for $m_t=174$ GeV. The same limits are 4.0 and 28 for $m_t=200$ GeV.

In the case where the $e^-$ beam is longitudinally polarized, we have
assumed the degree of polarization $P_e=\pm 0.5$, and determined 90\%
CL limits which can be achieved. In this case, the use of $P_e=+0.5$
and $P_e=-0.5$ is sufficient to constrain Im$c_d^{\g}$ and Im$c_d^Z$
simultaneously even though only one asymmetry (either ${\cal A}_{ch}$
or ${\cal A}_{fb}$) is determined. The 90\% CL bands corresponding to
$P_e=\pm0.5$ are shown in Figs. 2 and 3, for ${\cal A}_{ch}$ with
$\theta_0=60^\circ$, and for ${\cal A}_{fb}$ with
$\theta_0=10^\circ$, respectively. Again, these values of $\theta_0$
are chosen to maximize the sensitivity \cite{footnote}.

It can be seen from these figures that the simultaneous limits
expected to be obtained on Im$\cdg$ and Im$\cdz$ are, respectively,
about 0.45 and 1.5 for $m_t=174$ GeV from both the types of
asymmetries.  These limits are about 0.78 and 2.5 for $m_t=200$ GeV.
We see thus that the use of polarization leads to an improvement of
by a factor of about 5 in the sensitivity to the measurement of
Im$\cdg$, and by a factor of at least 10 in the case of Im$\cdz$.
Moreover, with polarization, either of ${\cal A}_{fb}$ and ${\cal
A}_{ch}$, with a suitably chosen cut-off, suffices to get the same
improvement in sensitivity.

Apart from simultaneous limits on Im$\cdgz$, we have also found out
the sensitivities of one of Im$\cdgz$, assuming the other to be zero,
using the $CP$-odd combination of angular distributions
$\f{d\sigma^+}{d\cos\theta} (\theta_l) - \f{d\sigma^-}{d\cos\theta}
(\pi-\theta_l)$ coming from eq. (19).  We assume that the data is
collected over bins in $\theta_l$, and add the 90\% CL limits
obtained from individual bins in inverse quadrature. We find that the
best individual limits are respectively 0.12 and 0.28 for Im$\cdg$
and Im$\cdz$, both in the case of $P_e=-0.5$, for $m_t=174$ GeV. The
corresponding limits for $m_t=200$ GeV are 0.18 and 0.43. As
expected, these limits are better than simultaneous ones. Even here,
there is an improvement due to polarization, but it is not as
dramatic as in the case of simultaneous limits.

Our limits on Im$\cdgz$ are summarized in Table 2.

\section{Conclusions}

We have calculated analytically the single-lepton angular
distribution in the production and subsequent decay of $\tt$ in the
presence of electric and weak dipole form factors of the top quark.
We have included effects of longitudinal beam polarization. We have
then obtained expressions for certain simple $CP$-violating angular
asymmetries, specially chosen so that they do not require the
reconstruction of the  $t$ or $\o t$ directions or energies. We have
analyzed these asymmetries to obtain simultaneous 90\% CL limits on
the imaginary parts of the electric and weak dipole couplings which
would be possible at future linear $\ee$ collider operating at
$\sqrt{s}= 500$ GeV and with a luminosity of 10 fb$^{-1}$. Figs. 1-3
show the allowed regions in the Im$\cdg$--Im$\cdz$ plane at the 90\%
CL. Table 2 summarizes the 90\% CL limits on Im$\cdgz$ in various
cases.

Our general conclusion is that the sensitivity to the measurement of
dipole couplings is improved considerably if the electron beam is
polarized, a situation which might easily obtain at linear colliders.
Another general observation is that the sensitivity is better for a
lower top mass than a higher one.

If we compare these results for sensitivities with those obtained in
\cite{PP}, where we studied asymmetries requiring the top momentum
determination, we find that while the sensitivities with the
asymmetries studied here are worse by a factor of about 3 in the
unpolarized case, the limits in the polarized case are higher by a
factor of about 2 as compared to the those in \cite{PP}. It is likely
that since in the experiments suggested here, only the lepton momenta
need be measured, improvement in  experimental accuracy can easily
compensate for these factors. A detailed simulation of experimental
conditions is needed to reach a definite conclusion on the exact
overall sensitivities.

We have also compared our results with those of \cite{cuypers}, where
CP-odd momentum correlations are studied in the presence of $e^-$
polarization.  With comparable parameters, the sensitivities we
obtain are comparable to those obtained in \cite{cuypers}.  In some
cases our sensitivities are slightly worse because we require either
$t$ or $\overline{t}$ to decay leptonically, leading to a reduced
event rate.  However, the better experimental efficiencies in lepton
momentum measurement may again compensate for this loss.

As mentioned earlier, since we consider only the electron beam to be
polarized, the asymmetries considered here can have backgrounds from
order-$\alpha$ collinear initial-state photon emission, which, in
principle,  have to be calculated and subtracted. However, in case of
correlations, it was found in \cite{back} that the background
contribution can be neglected for the luminosity we assume here. This
is likely to be the case in the asymmetries we consider here.

The theoretical predictions for $c_d^{\g,Z}$ are at the level of
$10^{-2}-10^{-3} $, as for example, in the Higgs-exchange and
supersymmetric models of CP violation \cite{bern,asymm,new}.  Hence
the measurements suggested here cannot exclude these modes at the
90\% C.L.  However, as simultaneous model-independent limits on both
$c_d^{Z}$ and $c_d^{\g}$, the ones obtainable from the experiments we
suggest, are an improvement over those obtainable from measurements
in unpolarized experiments.

Increase in polarization beyond $\pm 0.5$ can increase the
asymmetries in some cases we consider.  Also, a change in the
$e^+\,e^-$ cm energy also has an effect on the asymmetries.  However,
we have tried to give here only the salient features  of the outcome
of a possible experiment in the presence of longitudinal beam
polarization.

Inclusion of experimental detection efficiencies may change our
results somewhat.  However, the main thrust of our conclusions, that
longitudinal beam polarization improves the sensitivity, would still
be valid.
\newpage
\noindent{\Large \bf Appendix}
\vskip .5in

The expressions for $A_i$, $B_i$, $C_i$ and $D_i$ occurring in
equation (8) are listed below.

\begin{eqnarray}
A_0&=& \left\{2 (2-\beta^2) \left[ 2{c_v^{\g}}^2 +
2(r_L+r_R)c_v^{\g}c_v^Z + (r_L^2+r_R^2){c_v^Z}^2 \right] \right.
\nonumber \\
&& \left. + 2 \beta^2 (r_L^2+r_R^2){c_a^Z}^2 \right\} (1- P_e
P_{\overline e}) \nonumber \\ && + \left\{2 (2-\beta^2) \left[
2(r_L-r_R)c_v^{\g}c_v^Z + (r_L^2-r_R^2){c_v^Z}^2 \right] \right.
\nonumber \\
&& \left.  + 2 \beta^2 (r_L^2-r_R^2){c_a^Z}^2 \right\} (P_{\overline
e}-P_e), \nonumber \\ A_1&=& -8 \beta c_a^Z \left\{
\left[(r_L-r_R)c_v^{\g} + (r_L^2-r_R^2)c_v^Z \right]  (1-
P_eP_{\overline e}) \right. \nonumber \\ && \left. +
\left[(r_L+r_R)c_v^{\g} + (r_L^2+r_R^2)c_v^Z \right] (P_{\overline
e}-P_e) \right\},  \nonumber\\ A_2&=& 2 \beta^2 \left\{ \left[
2{c_v^{\g}}^2 + 2(r_L+r_R)c_v^{\g}c_v^Z + (r_L^2+r_R^2)\left(
{c_v^Z}^2 + {c_a^Z}^2 \right)  \right]  (1- P_e P_{\overline e})
\right. \nonumber \\ && \left. +  \left[ 2(r_L-r_R)c_v^{\g}c_v^Z +
(r_L^2-r_R^2)\left( {c_v^Z}^2 + {c_v^Z}^2 \right) \right]
(P_{\overline e}-P_e) \right\}, \nonumber\\ B_0^{\pm}&=& 4\beta
\left\{  \left(c_v^{\g} + r_Lc_v^Z\right) \left(r_L c_a^Z \mp {\rm
Im}c_d^{\g} \mp r_L {\rm Im}c_d^Z \right) (1-P_e)(1+P_{\overline e})
\right. \nonumber \\ && \left.+  \left(c_v^{\g} + r_Rc_v^Z\right)
\left(r_R c_a^Z \mp {\rm Im}c_d^{\g} \mp r_R {\rm Im}c_d^Z \right)
(1+P_e)(1-P_{\overline e}) \right\}, \nonumber \\ B_1&=& -4 \left\{
\left[(c_v^{\g}+r_Lc_v^Z)^2+ \beta^2 r_L^2 {c_a^Z}^2 \right]
(1-P_e)(1+P_{\overline e}) \right. \nonumber \\ && \left.-
\left[(c_v^{\g}+r_Rc_v^Z)^2+ \beta^2 r_R^2 {c_a^Z}^2 \right]
(1+P_e)(1-P_{\overline e}) \right\}, \nonumber \\ B_2^{\pm}&=& 4\beta
\left\{  \left(c_v^{\g} + r_Lc_v^Z\right) \left(r_L c_a^Z \pm {\rm
Im}c_d^{\g} \pm r_L {\rm Im}c_d^Z \right) (1-P_e)(1+P_{\overline e})
\right. \nonumber \\ && \left.+  \left(c_v^{\g} + r_Rc_v^Z\right)
\left(r_R c_a^Z \pm {\rm Im}c_d^{\g} \pm r_R {\rm Im}c_d^Z \right)
(1+P_e)(1-P_{\overline e}) \right\}, \nonumber\\ C_0^{\pm}&=&4
\left\{ \left[ (c_v^{\g} + r_Lc_v^Z)^2 \pm \beta^2
\gamma^2 c_a^Z \left( {\rm Im} c_d^{\g} r_L + {\rm Im} c_d^Z
r_L^2 \right) \right] (1-P_e)(1+P_{\overline e}) \right. \nonumber \\
&& \left. - \left[ (c_v^{\g} + r_Rc_v^Z)^2 \pm \beta^2
\gamma^2 c_a^Z \left( {\rm Im} c_d^{\g} r_R + {\rm Im} c_d^Z
r_R^2 \right) \right] (1+P_e)(1-P_{\overline e}) \right\}, \nonumber\\
C_1^{\pm}&=&- 4\beta \left\{  \left(c_v^{\g} +
r_Lc_v^Z\right) \left(r_L c_a^Z \pm \gamma^2 {\rm Im}c_d^{\g} \pm r_L
\gamma^2 {\rm Im}c_d^Z \right) (1-P_e)(1+P_{\overline e}) \right.
\nonumber \\ && \left.+  \left(c_v^{\g} + r_Rc_v^Z\right) \left(r_R
c_a^Z \pm \gamma^2 {\rm Im}c_d^{\g} \pm r_R
\gamma^2 {\rm Im}c_d^Z \right) (1+P_e)(1-P_{\overline e})
\right\},\nonumber \\
D_0^{\pm}&=& \mp 4 \beta \gamma^2 \left\{ \left(c_v^{\g} +
r_Lc_v^Z\right) \left( {\rm Re}c_d^{\g} + r_L {\rm Re}c_d^Z \right)
(1-P_e)(1+P_{\overline e}) \right. \nonumber \\ && \left.-
\left(c_v^{\g} + r_Rc_v^Z\right) \left({\rm Re}c_d^{\g} + r_R {\rm
Re}c_d^Z \right) (1+P_e)(1-P_{\overline e}) \right\}, \nonumber \\
D_1^{\pm}&=&\pm 4 \beta^2 c_a^Z \left\{ r_L \left( {\rm Re}c_d^{\g} +
r_L {\rm Re} c_d^Z \right) (1-P_e)(1+P_{\overline e})
\right. \nonumber \\
&& \left. + r_R \left( {\rm Re}c_d^{\g} + r_R {\rm Re} c_d^Z \right)
(1+P_e)(1-P_{\overline e})
\right\}. \nonumber
\end{eqnarray}

\newpage
\def \n {\noindent}
\def \v{\vskip .5cm}
\centerline{\Large \bf Figure Captions}
\vskip 1.5cm
\n {\bf Fig. 1.} Bands showing simultaneous 90\% CL limits on
Im $c_d^{\gamma}$ and
Im $c_d^Z$ using ${\cal A}_{fb}$ and ${\cal A}_{ch}$ with unpolarized
electron beam at cm energy 500 GeV and cut-off angle $35^{\circ}$.
Mass of the top quark is taken to be (a) 174 GeV and (b) 200 GeV.
\v

\n {\bf Fig. 2.} Bands showing simultaneous 90\% CL limits on
Im $c_d^{\gamma}$ and
Im $c_d^Z$ using ${\cal A}_{ch}$ with different beam polarizations,
and at a cm energy of 500 GeV and cut-off angle $60^{\circ}$.  Mass
of the top quark is taken to be (a) 174 GeV and (b) 200 GeV.
\v

\n {\bf Fig. 3.} Bands showing simultaneous 90\% CL limits on
Im $c_d^{\gamma}$ and
Im $c_d^Z$ using ${\cal A}_{fb}$ with different beam polarizations,
and at a cm energy of 500 GeV and cut-off angle $10^{\circ}$. Mass
of the top quark is taken to be (a) 174 GeV and (b) 200 GeV.

%%%%%%%%%%%%%%%%%%%%%%%%%%%%%%%%%%%%%%%%%%%%%%%%%%%%%%%%%%%%%%%%%%%
%%%%%%%%%%%%%%%%%%
%%%%%%%%%%%%%%%%%%     captions
%%%%%%%%%%%%%%%%%%
%%%%%%%%%%%%%%%%%%%%%%%%%%%%%%%%%%%%%%%%%%%%%%%%%%%%%%%%%%%%%%%%%%%
\def \v{\vskip .5cm}
\newpage
\centerline{\Large \bf Table Captions}
\vskip 1.5cm
\n {\bf Table 1.} Number of $t \bar t$ events, with either \t  or
\tbar
decaying leptonically,
for c.m. energy 500 GeV and integrated luminosity $10\;{\rm fb}^{-1}$
for two different top masses with polarized and unpolarized electron
beams at different cut-off angles $\theta_0$.
\v
\n{\bf Table 2.}  Limits on dipole couplings obtainable from different
asymmetries. In case (a) limits are obtained from ${\cal A}_{ch}$ and
${\cal A}_{fb}$ using unpolarized beams (Fig. 1), and in case (b)
from either of ${\cal A}_{ch}$ (Fig. 2) and ${\cal A}_{fb}$ (Fig. 3)
with polarizations $P_e=0,\;\pm 0.5$.  Charge-asymmetric angular
distribution is used in case (c) where $0\;{\rm and}\;\pm 0.5$
polarizations are considered separately.  All the limits are at 90\%
CL.

\newpage
\begin{center}
\begin{tabular}{||c|c|c|c||c|c|c||}
\hline
&\multicolumn{3}{c||}{$m_t=174$ GeV}&\multicolumn{3}{c||}{$m_t=200$
GeV}\\
$\theta_0$& $P_e=-0.5$&
$P_e=~0$&$P_e=+0.5$&$P_e=-0.5$&$P_e=~0$&$P_e=+0.5$
\\
\hline
$0^\circ$&1003&845&687 &862&723&585\\
$10^\circ$&988&832&675 &849&712&576\\
$35^\circ$&826&689&553 &711&593&475\\
$60^\circ$&507&419&330 &438&362&286\\
\hline
\multicolumn{7}{c}{}\\
\multicolumn{7}{c}{Table 1}\\
\multicolumn{7}{c}{}
\end{tabular}
\vskip 2in
%%%%%%%
%%%%%%% Table 2.
%%%%%%%
\begin{tabular}{|ll|c|c|c|c|}
\hline
\multicolumn{2}{|l|}{}&\multicolumn{2}{c}{$m_t=174$ GeV}&
\multicolumn{2}{c|}{$m_t=200$ GeV}\\[3mm]
\multicolumn{2}{|l|}{Case}&$|{\rm Im}c_d^{\gamma}|$&$|$Im$
c_d^Z|$&$|$Im$c_d^{\gamma}|$&$|$Im$c_d^Z|$\\[2mm]
\hline
(a)  unpolarized&&2.4\hskip 5mm&17&  4.0&28\\[3mm]
(b)  polarized($P_e= 0,\;\pm$0.5)&&0.45&1.5&0.78&2.5\\[3mm]
(c)  angular distribution:&$P_e=+0.5$&0.13&0.74&0.21&1.21 \\
&$P_e=~~0.0$&0.13&0.81&0.20&1.30\\
&$P_e=-0.5$&0.12&0.28&0.18&0.43\\[2mm]
\hline
\multicolumn{6}{c}{}\\
\multicolumn{6}{c}{Table 2}\\
\multicolumn{6}{c}{}
\end{tabular}
%%%%%%%%%%%%%%%%%%%%%%%%%%%%%%%%      end    %%%%%%%%%%%%%%%%%%%%%%%%
\end{center}

\begin{thebibliography}{99}
\bibitem{footnote} In case of ${\cal A}_{fb}$, the choice
$\theta_0=0$ also gives very similar sensitivities. However, since
this condition would be impossible to achieve in a practical
situation, we choose a nonzero value of $\theta_0$.
\end{thebibliography}
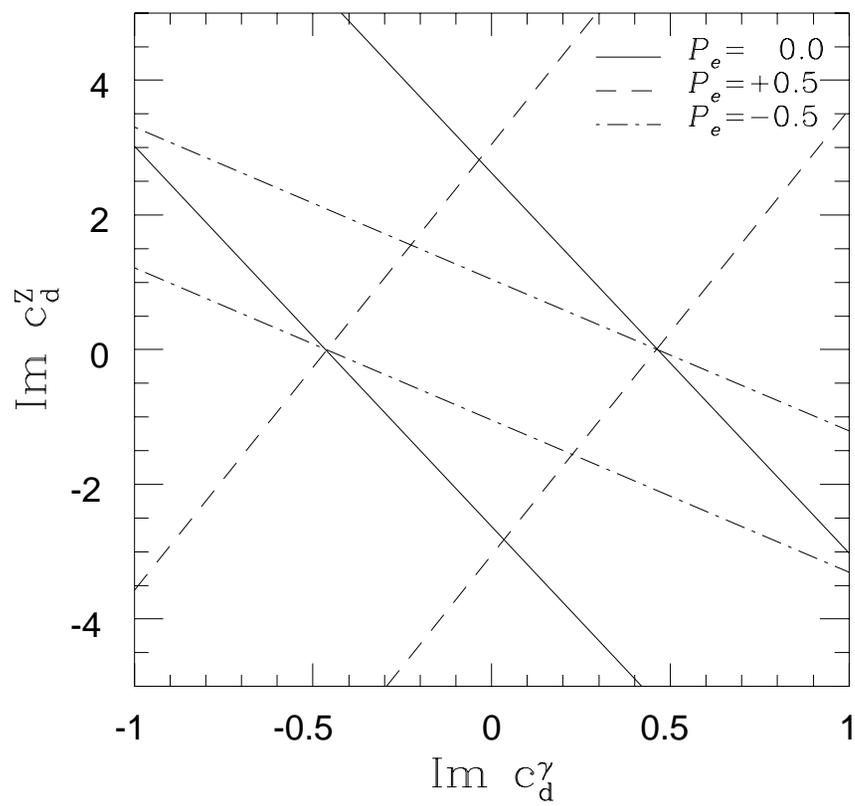
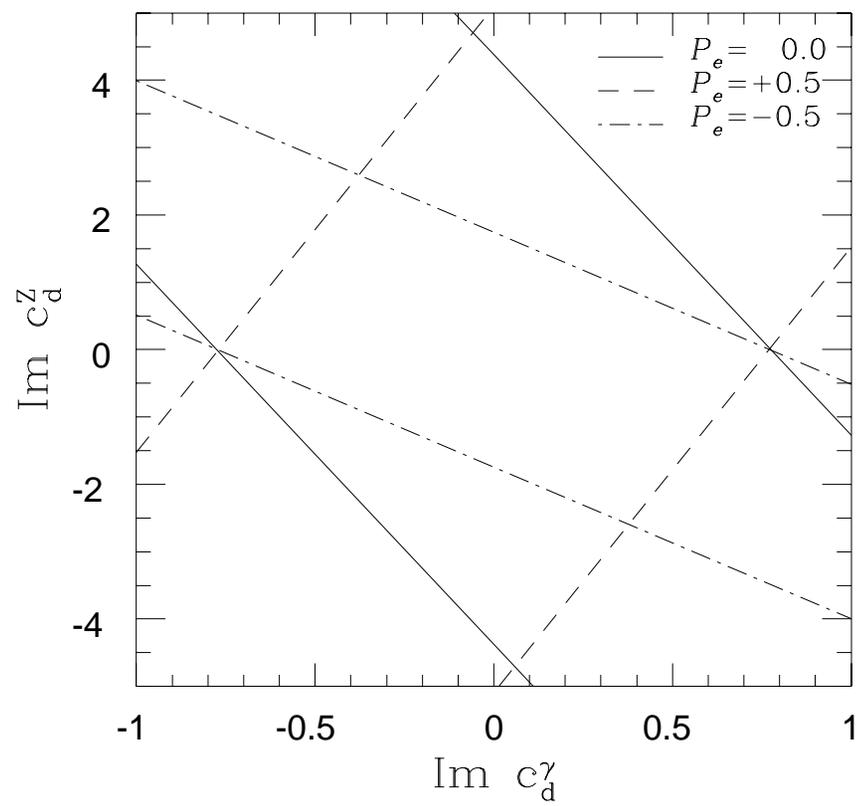
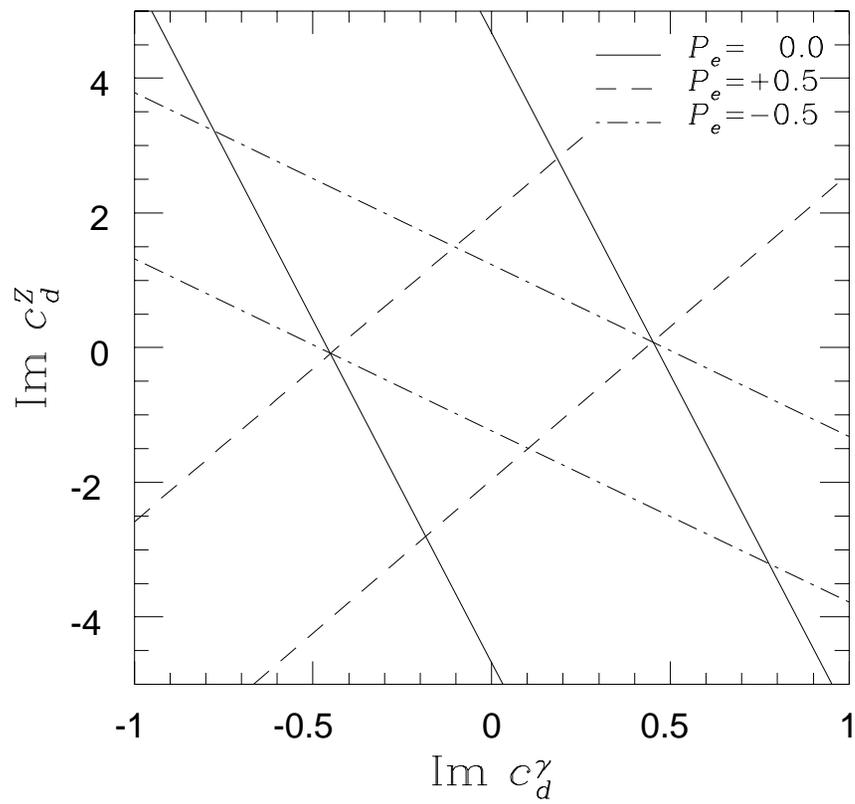
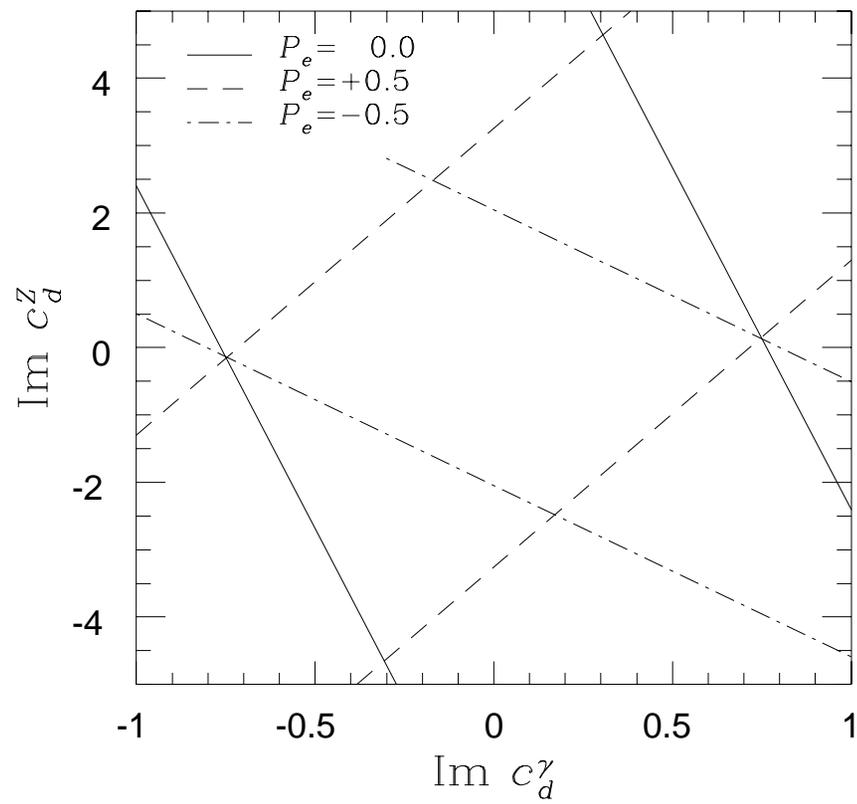
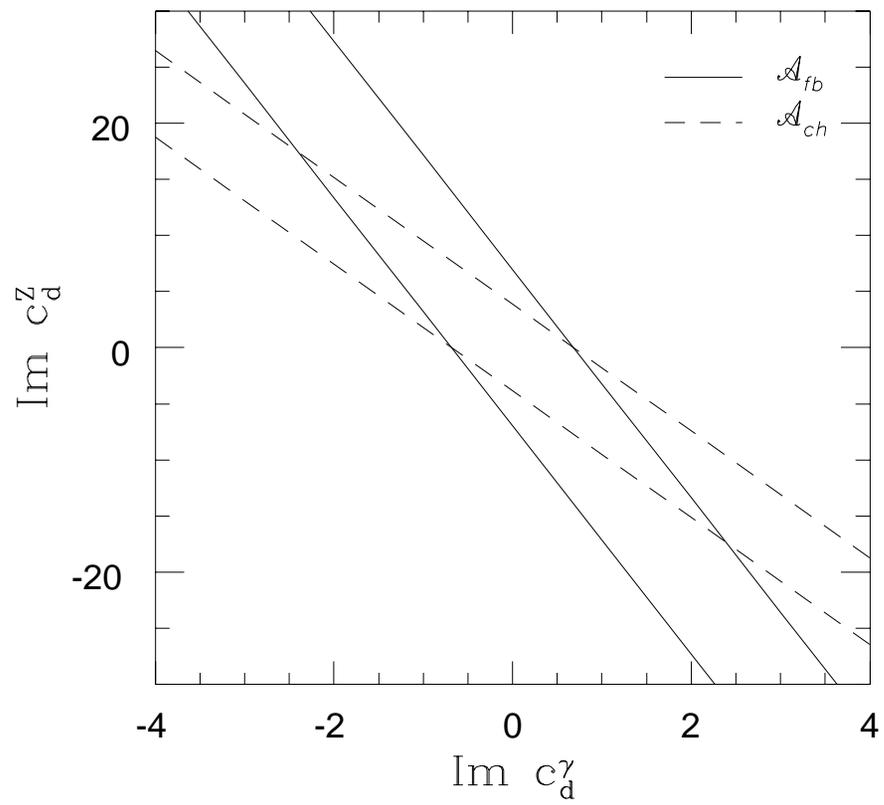
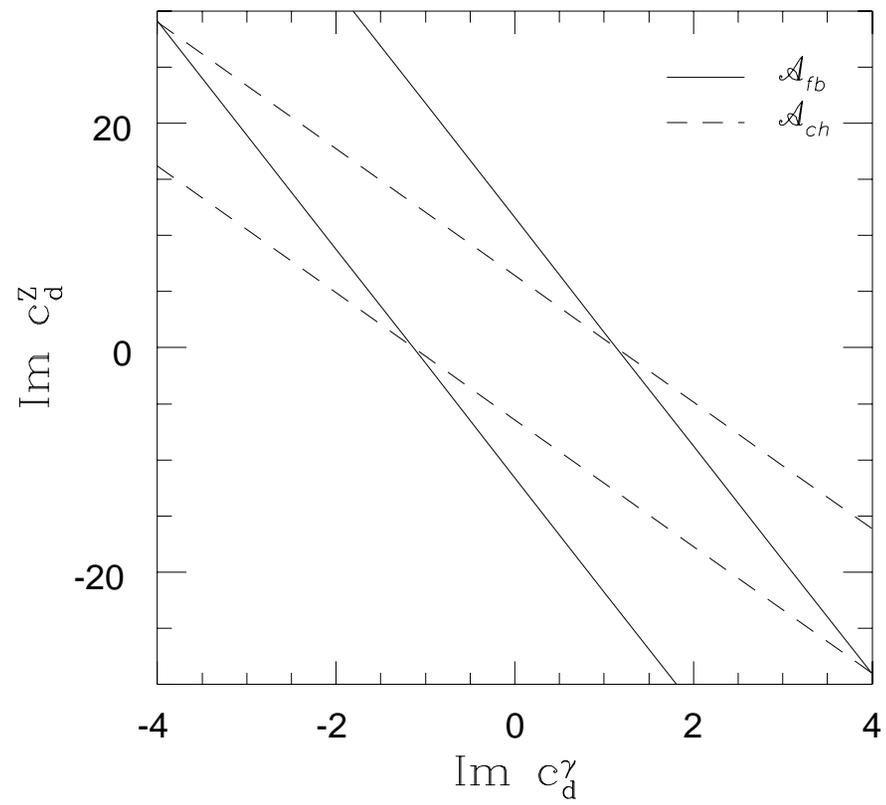
\end{document}